\documentclass[fleqn]{article}

\newcommand{\titel}
{Questions related to Bitcoin and other Informational Money}

\usepackage{stmaryrd,amssymb,amsmath,amsthm,url}

\theoremstyle{definition}

\newcommand{\Nat}{{\mathbb N}}

\title{\titel}

\author{
	Jan A.\ Bergstra \& Karl de Leeuw
	 \\
\\
  {\small
	  Informatics Institute,
	  University of Amsterdam}\\
	{\small Email: \url{j.a.bergstra@uva.nl}, \url{karl.de.leeuw@xs4all.nl}
	}
}
\date{}

\begin{document}
\maketitle

\begin{abstract}
A collection of questions about Bitcoin and its hypothetical relatives Bitguilder
and Bitpenny is formulated. These questions concern 
technical issues about protocols, security issues, issues about the formalizations of informational monies
in various contexts, and issues about forms of use and misuse. Some questions are
formulated in the more general setting of informational monies and near-monies.
 
We also formulate questions about legal, psychological, and ethical aspects of informational money.
Finally we formulate a number of questions concerning the economical merits of and outlooks for Bitcoin. 
\\[5mm]
\emph{Keywords and phrases:}
informational money,
exclusively informational money,
Bitcoin, Bitguilder, Bitpenny, interest prohibition, gambling prohibition.
\end{abstract}

\newpage
{\small\tableofcontents}~\newline~\newline

\setcounter{section}{-1}
\section{Introduction}

\noindent In this paper we\footnote{%
Disclosure: using the terminology of  \cite{BergstraL2013}, at the time of writing the 
first author (thinks of himself that he)  ``owns'' 3.40050000 BTC (which were acquired for about 100 Euro
in total including bank transfer cost), as an indirect user of two participation services. 
At the time of submitting this paper to \texttt{arXiv}
both services are out of action reportedly due to various legal problems in need of urgent resolution.}
will proceed along the path chosen by the first author in \cite{Bergstra2013a}  and subsequently in joint work in \cite{BergstraL2013} 
where money was analyzed and defined, both in general terms, and more specifically in its modern manifestation 
of  informational money\footnote{%
In~\cite{Bergstra2012d} the use of the 
term informaticology (IY) has been advocated. Informaticology (IY) is decomposed as: 
IY = CS + DS + FS = Computer Science + Data Science + Fiction Science. Informational money is rooted
in each of these components: CS for protocols, encryption, computing, security, networking; DS for
analysis on distributed data sets resulting from transaction logging; FS for the computer game outlook
that successful informational monies must probably display. This paper is meant as a 
a contribution to the informaticology of informational money.}
with Bitcoin as a case study. Bitcoin was studied while making use of
available sources about Bitcoin only instead of inspecting versions of its open source client code. The latter method admittedly 
can be employed to find answers to some questions about Bitcioin, but it is less helpful for developing levels 
of abstraction that allow a theoretical 
study of the underlying issues. Our objective is to formulate a number of questions about informational monies,
and about Bitcoin and the related thought experiment Bitguilder that was put forward in \cite{BergstraL2013}. 
Below we will introduce
Bitpenny as a second hypothetical modification of Bitcoin geared towards investigating matters of 
usage and usage modeling 
from an abstract perspective. 

By collecting questions rather than answers we intend to draw a larger picture of a subject 
that may grow around informational monies 
thereby promoting a better  understanding of
the range of possible implementations of what we have called the Nakamoto Architecture in
\cite{BergstraL2013}.  in particular, 
and to a systematic process for improving the understanding of informational monies in general.

In writing this paper we run the risk being too specific by portraying Bitcoin 
as being more important than it deserves when compared to its
steadily growing family of relatives and descendants which may play competing as well as 
complementary roles.\footnote{%
A survey of such systems can be found in \cite{Steadman2013}.}
 Bitcoin-like systems are 
comparable to program notations or to operating systems in that there seems to be no end to the number of meaningful 
variations that can be contemplated. This paper is not the first one of its kind. In \cite{Barber2012} a range of questions
about Bitcoin is raised starting from a question that is still very much in the air: 
``Does Bitcoin have what it takes to become a serious candidate for a long-lived stable currency, 
or is it yet another transient fad?'' Except for this motivating question 
we will try not to repeat the issues put forward in that paper below.

\subsection{Taking Bitcoin seriously}
Whoever studies Bitcoin and other informational money in detail and starts writing about it takes significant risks. 
At any moment the designer of Bitcoin can be discovered or make him/herself known
with an unpredictable impact on how Bitcoin is perceived. At any moment regulators from various jurisdictions
may take action against Bitcoin and its participants with an unpredictable impact on the 
reputation of anyone who has been writing about it. At any moment the Bitcoin market may collapse due to changing sentiments
among Bitcoin participants. And of course at any moment technical problems may
bring Bitcoin down and without any advance warnings a technically stronger successor technology may appear on 
the scene and instantly degrade Bitcoin from being a prospect to being a mere historic incident. Authors writing 
about conventional monies face comparable risks but to a far lesser extent. 

Our perspective on these risks is that Bitcoin offers a novel approach to informational money which merits the attention of 
academic research. Undeniably it is an advantage for an academic author on Bitcoin that it is a live system.
However, after its demise that is to be expected sooner or later, we expect that the architecture of Bitcoin will still stand 
out as a significant
milestone in the evolution of informational monies. 

Not writing about Bitcoin and other novel informational monies also involves risks which can be 
contrasted with the risks just mentioned. The manifest risk of staying away from these new developments  
is that one needlessly failed to try to contribute to a very significant
development at a time that it might have been most rewarding.
 
\section{Questions about software technology and mining}
A reasonable perspective on Bitcoin is to view it as a steadily evolving piece of software. 
That perpective gives rise to a focus on protocols, algorithms, security models and the like. 
\subsection{Bitcoin protocol}
Regular reference is made to the so-called Bitcoin protocol. That leads us to formulating several questions.
\subsubsection{What is the Bitcoin protocol?}
 Probably some abstraction of the 
Bitcoin open source client code is meant. We could not find decisive information about the Bitcoin protocol (a finding
previously  reported in~\cite{BabaioffDOZ2012}), and we 
are led to the hypothesis that such a protocol, in the form of an agreed upon abstraction of existing 
software which imposes constraints on other of future implementations, does not exist. 
So the question is: is there a Bitcoin protocol, and if
so, what is it?\footnote{%
On \texttt{https://en.bitcoin.it/wiki/Protocol\_specification} a Bitcoin protocol specification is provided. We feel that
a more abstract and rigorous specification is needed which explicitly clarifies the degrees of freedom
of implementation allowed for by the protocol. In~\cite{Boehme2013} it is argued that a useful protocol description 
must take care of the success factors listed in RFC 5218 (see~\cite{ThalerA2008}), 
while some potential success factors such as the purported but unproven specific 
usability (of Bitcoin) for fraudulent purposes must not make it into an RFC style protocol specification. For recent
information on abuses of virtual currencies see~\cite{Huang2013}.}
\subsubsection{When, if ever,  will an open protocol specification become leading?}
It would be simpler to appreciate the Bitcoin development for an outsider if instead of a 
client a protocol is the objective of open development. Is this a likely thing to happen and if so, how will that work?
\subsubsection{Find Bitpenny?}
\label{bitpenny}
Bitpenny (very simplified Bitcoin) is a name that we ``coin''  for a protocol, or rather process specification, that provides 
the behavior of Bitcoin from the viewpoint of non-mining users only. Bitpenny takes care of validation and new coin creation through abstract mechanisms in need of further distributed 
implementation. Bitcoin might be considered an implementation of Bitpenny.

Because of its abstraction level Bitpenny does not, and need not, 
feature the phenomenon of forking and subsequent recovery from forking. Bitpenny prevents 
double-spending ``by magic''.
We claim that an appropriate design of Bitpenny will constitute an adequate point of departure
for studying in a rigorous and formal manner the behavior of a Bitcoin-like system from a user perspective. 

The question is to determine an appropriate Bitpenny specification that meets the above requirements and provides
a useful carrier for theoretical investigation of its usage.

\subsection{Bitcoin algorithms}
Algorithms are equivalence classes of implementations of functional specifications. More specifically 
algorithms consist of architectural ideas (in principle amenable for patenting) about ways of implementing
functionalities. An algorithm can be equated abstractly with its extension, 
that extension being the a family of classes of its realizations in a suitably corresponding family of program notations.
For a recent survey on conceptions of the notion of an ``algorithm" we refer to~\cite{BergstraM2013}.

Users of Bitcoin need only understand functionalities, but in order to grasp how it works some intuition concerning
algorithms for the implementation of such functionalities is probably necessary. 
In principle, however, we adhere to the view that Bitcoin is
determined by an (emerging) functional specification (interface or protocol) for which increasingly performant algoritmns 
are invented (or must on say discovered) and for which corresponding implementations are developed with
increasing (non-functional and non-performance related) quality characteristics.
\subsubsection{How to clarify the algorithmic content of Bitcoin for its participants?}
The complexity of Bitcoin cannot be talked away by the many blogs and short articles that suggest giving an
explanation of it.  It ought to be an objective for Bitcoin development that what is produced is amenable to some form
of abstract understanding allowing a participant to understand precisely how transfers work, what miners are doing, 
how transactions are validated, how the blockchain is stored and so on.

The comprehensibility of the system should as much be a design objective as its 
quality of not having a single point of failure or not being vulnerable for inflation emerging 
from political pressures. In \cite{BergstraL2013} an attempt has been made to understand Bitcoin in its 
quality of being an implementation of what is called a Nakamoto Architecture. Unfortunately that 
architecture falls short of constituting an abstraction of Bitcoin that is sufficiently informative for its users 
to be able to claim an understanding of what is going on. The question is how this state of affairs can be improved.

\subsubsection{Explanation of Bitcoin on the basis of simplified models of software}
In \cite{BM2009} the notion of control code has been defined with reasonable rigor. 
Control code shares with programs that
it can be used to control machine behavior, but it need not allow the mechanical interpretation of machine
behavior as the effectuation of an instruction sequence. 
Control code and its interaction with code controlled machines
can be viewed as an attempt to provide a minimalistic theory of computer software, in particular abstracting from
the idea that software encodes or represents algorithms. Several questions can be posed:
\begin{enumerate}
\item To what extent can a meaningful and complete explanation 
of Bitcoin be provided on the basis of code controlled machines but without making reference to the notion
of a computer program or to computer programming. 

\item If the control code fragments for Bitcoin can be represented as instruction sequences in the style of \cite{BL02}
(and for that reason can be considered programs), is that a useful 
extension of the expressive power of a theory of computer software (for this particle application).

\item In \cite{BM07} strategic interleaving has been proposed as a theory from first principles of multi-threading,
providing a simplified model that may be helpful in some but not in all cases. 
Is multi-threading in the form of strategic interleaving sufficiently expressive to understand the 
concurrency aspects of Bitcoin? (Or can Bitcoin be 
explained without any mention of concurrency?)
\end{enumerate}

\subsubsection{Real time limitations?}
Does the Bitcoin protocol and its dependence on globally maintaining the entire blockchain by many independent
participants  impose
limitations on the real time performance of current as well as future Bitcoin clients, or is there ample room
 for performance improvement within the degrees of freedom allowed by  the current protocol? 
 (See~\cite{Bruce2013} for a proposal on how to reduce the memory cost of blockchain storage.)

\subsubsection{Algorithmic complexity analysis of the proof of work mechanism}
\label{complexitySHA}
Here is a simplified version of the combinatorial problem that  miners are expected to solve in order to satisfy their
proof of work obligation. A recent description, and critique of Bitcoin mining can be found
in~\cite{CourtoisGN2013}. Of course that work can be done by a program running on a general 
purpose machine but we preferably think of a dedicated machine for solving the combinatorial puzzle.

In particular we imagine machines $M$ that compute a function $f$ from a finite domain
$D$ to a finite range $R$, where $f$ is implicitly defined by the constraint that for each $d \in D$, the result $r=f(d)$ 
satisfies some criterion $\phi(r,d)$.  Thus for $f$ we have an implicit definition that is expected to be satisfied 
by many different implementations.

We assume that provider $P_m$ supplies these machines and delivers the good (say machine $M$) 
in a box equipped with the following promises: 
\begin{enumerate}
\item the expected time of computing $f(d)$ is below $E_t(d)$, 
\item the distribution of running times is normal, (iii) the spread is $s$, and
\item the energy consumption when running is below $e$.
\end{enumerate}

Let $2^{n}$ represent
the set of bit sequences of length $n$. SHA-256-512 is the secure hash algorithm SHA-256 applied to 
$2^{512}$ (see~\cite{FIPS2008} for the Standard and~\cite{BergstraM2013} for a recent formalization
in terms of instruction sequences). We will abbreviate 
SHA-256-512 to $h$. As a domain $D$ we choose
$D = 2^{256} \times 2^{256}$, and the range $R$ is $2^{256}$. The criterion $\phi(d) = \phi(d_1,d_2)$ works as follows:
\[\phi(d_1,d_2,r) \Leftrightarrow \texttt{bs2n}(h(d_1 \mid \mid r)) \leq \texttt{bs2n}(d_2). \]

Here $\mid\mid$ denotes string concatenation and $\texttt{bs2n}$ converts bit sequences into natural numbers assuming
big Endian representation.\footnote{%
ASIC based machines that compute functions $f$ according to these concrete specifications, 
modulo some details, are now on the market as so-called Bitcoin mining tools.
At the time of writing Avalon and Butterfly are important Bitcoin mining tools.}

SHA-256-512 is presented by FIPS (see \cite{FIPS2008}) as a secure hashing function 
which is for that reason supposed to be 
resistant against so-called
pre-image attacks: given $r \in 2^{256}$ it is very difficult to find a string $d \in 2^{512}$ with $h(d) = r$. The statement that this is a 
difficult combinatorial problem, when measured in terms of the number of steps needed by a deterministic machine to obtain a solution, 
is a mathematical statement in principle but no proof is known. 

After some simplification and abstraction one might hold that 
Nakamoto's design of Bitcoin manufactures from SHA-256-512 a parametrized combinatorial problem. 
That problem is then used as the Bitcoin proof of work problem used for competitive transaction validation and rewarding its
winners with a mining yield: 
computing $f(d_1, d_2)$,
necessarily fast in order to win a competition and preferably at low energy costs on cheap but dedicated 
machines in order to make a profit as well.

This problem is assumed to be difficult to solve, and its difficulty is assumed to become 
exponentially harder with a linear growth of  the number of leading zeroes of $d_2$. 

The theoretical question that we pose is to prove or at least further consolidate the belief that this 
mining problem is hard, and practically intractable.

\subsubsection{Knowledge extraction from Bitcoin mining}
Is it possible to extract knowledge from the Bitcoin mining process that eventually leads to an effective
cryptanalysis of SHA-256-512. A successful cryptanalysis can be expected only under the assumption that 
this hash function is not secure on the long run.

Stated differently and following \cite{BergstraL2013}, can mining be turned into a multi-player 
game for breaking SHA-256-512?

\subsection{Nakamoto architecture and reference framework}
Whereas the Nakamoto architecture is meant to capture an abstraction of Bitcoin, a reference framework should capture
a common abstraction for a range of P2P informational monies and near-monies. 
\subsubsection{Reference framework for Bitcoin and Bitcoin alternatives}
The question is how to find an abstraction level at which Bitcoin can be compared with existing alternatives (e.g. 
Litecoin, Zerocoin, Freicoin)\footnote{%
See \texttt{http://en.wikipedia.org/wiki/List\_of\_cryptocurrencies} for a survey of so-called 
cryptocurrencies.},
future Bitcoin alternatives, and proposed modifications of Bitcoin. 
The Nakamoto Architecture of \cite{BergstraL2013} may serve as a starting point for defining workable abstractions.

\subsubsection{Surveying defenses against double-spending}
Robustness against double-spending attacks (see e.g. \cite{Herrmann2012}) in a P2P system without single 
point of failure is the key technical problem that justifies the introduction of mining. 
Is it possible to give a survey of methods  that might be used to defend a P2P informational money 
against double-spending attacks, and preferably to provide a mathematically meaningful classification 
of such methods? As a motivation for this issue we notice that 
only with such
a survey at hand the rationale for the algorithmic complexity and the  induced computational cost of Bitcoin can be assessed.

\subsubsection{Under which circumstances is ECDSA a limiting factor for Bitcoin?}
Digital signatures using elliptic curve cryptography (see ECDSA in \cite{JohnsonMV2001} and 
further see~\cite{BosEA2013}) 
constitute an essential ingredient for Bitcoin.
The question is to analyze in qualitative as well as in quantitive terms under which conditions ECDSA becomes a limiting
factor for Bitcoin and how an upgrade of that part of the Bitcoin security model might work. 

The required analysis must preferably be so
clear and convincing that for an agent contemplating investment of say EUR in Bitcoin, refraining from that investment 
because of fear that ECDSA weakness will bring Bitcoin on its knees becomes implausible.

\subsection{Conceptual analysis of anonymity}
A reference architecture for informational monies is also needed for the development of a 
conceptual framework concerning anonymity for Bitcoin and comparable informational monies or near-monies. 
In spite of a number of works on Bitcoin and anonymity we feel that a clear conceptual analysis of 
what anonymity means in the case of informational monies 
and Bitcoin-like P2P realizations thereof is still largely missing.

\subsubsection{Bitcoin formalization with informational identities for its participants}
A convincing explanation of concepts of anonymity can be found in~\cite{HalpernO2005}. An immediate conclusion
one may draw from that paper is that without taking informational identities (IIDs) of participants on board 
it is theoretically impossible to investigate anonymity in the context of Bitcoin. 
The problem is to develop a model, say Bitcoin-IID,
that can serve as the basis of a theory of anonymity for the practice of Bitcoin.

\subsubsection{Develop a theory of anonymity for Bitcoin-IID}
Many papers have already been written about issues in connection with Bitcoin anonymity 
(see e.g. \cite{AndroulakiKRSC2012}). The problem is to develop a comprehensive theory 
of anonymity for Bitcoin-IID 
starting with fundamental principles such as can be found in~\cite{HalpernO2005} or 
in~\cite{HughesS2004} and based 
on a systematic terminology comparable to that of~\cite{PfitzmannK2001}.

\subsubsection{What is pseudonimity}
In~\cite{BritoC2013} pseudonimity rather than anonymity is claimed to describe what Bitcoin has
on offer in terms of not revealing user names. The question is to define pseudonimity in a
mathematically precise and quantifiable manner (as has been done for anonymity) and to apply that definition
to the particular case of Bitcoin.\footnote{%
In~\cite{PfitzmannK2001} pseudonimity is merely a qualitative notion,
which is somehow less convincing.}

\subsection{Wallets and keys}
A wallet is described as a secure and password protected 
software tool with built-in backup facility that is able (i.e. enables a user) to safely
maintain a collection of addresses with corresponding private keys. (For a remarkably
informative technical survey of Bitcoin including wallets see~\cite{Entrup2013}.)

The wallet operates in the context of a client which is equipped with software features 
enabling its user to issue outgoing transactions, to confirm the success
of preceding outgoing transactions by inspecting the stream of incoming blocks, and to
participate in peer voting in order to collectively determine the growth of the blockchain with 
other peers in the network.
\subsubsection{Options for address removal}
Assume that A is in control of address $k$ (that is, A has control over $k$) 
and that A is in control of its corresponding secret key $s$, and that A intends not to use $k$ 
anymore indefinitely (for instance out fear that the secret key has leaked). Suppose that A intends to avoid
other agents to make transfers to that account but that at the same time A is not in the position to 
communicate that fact to relevant peers. This setting leads to several questions:

(i) How should A proceed in the case of Bitcoin (i.e. $k$ is a Bitcoin address).

(ii) Can Bitcoin be extended with the following {\em address blocking service}: A can
A actively destroy $k$ (with the authority of digitally signing with $s$) so that 
(a) no future incoming transfers to $k$ will be enabled from whatever source, and (b) A can safely forget about $s$
and more importantly, (c) A can abandon the non-trivial task of its secure storage. A difficulty is
what fee A ought to pay for the blocking service as an incentive for incorporating a sucessful event of 
address blocking in the blockchain.\footnote{%
Designing incentives in P2P systems has been investigated in detail in~\cite{Berciu2013}.}

(iii) Is a temporary blocking service useful, and if so, how can it be designed and implemented.

\subsubsection{Classification of participation services for wallets and exchange}
Many agents nowadays so-called participation services for Bitcoin in the terminology of \cite{BergstraL2013}. 
Given the growth in number and diversity of participation services developing a
classification of participation services becomes a necessity, together with a specialized terminology. 

Assuming that an agent running a Bitcoin client has access (or exclusive control) to addresses and corresponding secret keys, 
what derived forms of control can a participation service offer to its customers.

The simplest idea, at least in theory, is that an agent outsources the operation of a 
Bitcoin client to a service provider, in which case the client still
has access to and control over public and corresponding secret keys, but with the software for handling those
instruments taken from the provider. The question (or rather problem) 
is to develop a useful classification of services that can be
delivered by Bitcoin participation services.

\subsubsection{Formalization of wallets}
Users of Bitcoin can hardly escape making use of a wallet. Clear and rigorous, though sufficiently abstract,
specifications of wallets are needed, and so is some form of classification of different functional options for wallets.
\section{Use and misuse}
The use and misuse of (exclusively) informational monies differs in part from that of ordinary monies. 
We try to capture these 
differences in a few questions that need some extensive preparations.
\subsection{Use}
Use concerns those interactions with a system that explain the existence of a system in the first place.
\subsubsection{Is participant authentication really unneccesary?}
A remarkable property of Bitcoin is that it provides no support for identity management and authentication of agents
who act as payers, payees, and miners. It seems obvious that once Bitcoin is used at a larger scale such mechanisms must
be included as ``add on features" at least. The question that can be posed is to what extent the core of a large
scale financial transaction system can indeed do without agent identity management and corresponding authentication.
\subsubsection{Payer (and payee) authentication}
If agent A who is in control of account $k$ plans to transfer $q$ BTC to agent B who claims to be in control of account $l$, then
at some stage B may need to make sure that it was A who paid and not some other agent C who was
actually in control of $k$. B may need to prevent A from making use of its knowledge that C plans and performs
an equivalent payment. How in principle should B (and perhaps also A) go about this kind of issue?

B may ask A, after a transfer from account $k$ has been received on account $l$ (which B knows to be under
its own control) to return a challenge, say $c$ (a bit sequence randomly produced by B), with a digital signature
that can be checked via key $k$. If A responds correctly and quickly to this request, B knows that A was in 
control of $k$ at some moment after said transfer from $k$ to $l$ was issued and validated. At this point some convention is needed, for instance that A only signs challenges proving its control over $k$ if $A$ agrees that it has been issuing all
preceding transactions taking amounts from $k$, that is A has not witnessed any activity of other agents 
who must be or must have been in control of account $k$ as well in order to perform that particular activity.

It appears that complete and reliable transaction logging is essential for a Bitcoin user. If so that implies that
wallet functionality is a necessity. Is this conclusion valid?

\subsubsection{Decision taking process for Bitcoin transfers}
An agent who plans to transfer an amount of BTC to a another agent needs to perform some decision taking
that takes place as a subprocess of a more comprehensive decision making process.
The question is to what extent generic process architectures for decision taking and decision making can be developed,
which are tailored to the specific context of Bitcoin related activities.\footnote{%
In \cite{Bergstra2012} we have specified a decision taking process for initiating a thread  which takes care of all steps
that must be performed when a valuable object is to be sold.}

\subsubsection{Decision taking for Bitpenny transfers}
Bitpenny (see \ref{bitpenny} above) is a (hypothetical) theoretical abstraction of Bitcoin abstracting from 
the phenomenon of double spending attacks and by therefore from the mechanics
of mining. Bitpenny is supposed to be more amenable to theoretical analysis than Bitcoin is. 
The question is to develop a theory of decision taking for Bitpenny transfers. 

\subsubsection{Potential unintended transfer of excessive fees}
Consider an organization OABD (Organization Accepting Bitcoin Donations)
 that has publicly announced that it will accept donations made in BTC on an
account (=public key), say $k$,  which is to be found on their website in hexadecimal notation.\footnote{%
Wikileaks is a well-known example for that state of affairs.}

Every now and then OABD will wish to transfer an amount from $k$ where
donations are collected to say $l$ from which they intend to perform payments 
(assuming that they intend more than merely hoarding donated BTCs).

Suppose they transfer as follows:
\begin{enumerate}
\item assume that $s$ is the secret key belonging to $k$ and that $s$ is only accessible to OABD,
\item assume that an amount $q = q(k)$ has been collected on $k$ (this value OABD 
retrieves from the blockchain, which is publicly available),
\item OABD intends to transfer $r$ from $k$ to $l$ (assuming that $r+f = q$), where
\item $f$ is the fee that OABD is willing to allow miners to collect when validating this transfer,
\item OABD places the transfer instruction \texttt{sign}($s$,($k,r,l$)) on the 
P2P network in order to be noticed by all participants, 
including miners who quickly start validating this transaction for inclusion in a new block.

\item this will have the expected effect (in principle, and irreversible for OABD) that (i) $r$ is transferred to $l$, 
(ii) (assuming that $k \neq l$) $k$ ends up at ``balance'' zero, and (iii) an additional amount $f$ is now in the 
hands of a successful miner.
\end{enumerate}

Now suppose that immediately after the transfer has been placed, an incoming transfer (probably a donation) to 
$k$ with amount $g$ is validated (that validation taking place before validation of the above outgoing transfer): 
then the successful miner (who mines OABD's outgoing internal from $k$ to $l$) will earn $f+g$ 
(and the amount $g$ seems to be lost for OABD).

The question is: what protects Bitcoin donation receiving participants 
against this problem. 
If the problems is real, then how should multiple donation expecting and receiving participants act in 
order to prevent this kind of unfortunate course of events.

\subsection{Access related misuse, illegal use, and incidentally legalized use}
Drawing a line between use and misuse is difficult, and probably unnecessary. 
It is practical to avail of a terminology in which
both use and misuse can be described impartially with an open eye for the fact that what constitutes use today 
may be labeled as misuse tomorrow and conversely. Seen from our perspective this unclarity 
 introduces a bias against the use of terms like theft, malicious code, 
hacking, attack etc. Each of these terms share  negative connotations to an extent 
that they are rendered useless as base concepts
and must be defined as derived concepts instead.

\subsubsection{Can we find a general terminology for use and misuse?}
In \cite{BergstraL2013} we have coined Bitguilder as a clone of Bitcoin with different rules of engagement. In 
particular in Bitguilder there is no concept of ownership except access. 
As a consequence there is no notion of theft in Bitguilder. For that reason the term theft cannot be used to explain 
misuse in Bitguilder 
and by implication ``theft'' cannot be accepted as  a constituent of a general 
explanation of misuse for the full range of informational monies. 
The question is how to develop a general terminology for use and misuse.

\subsubsection{Classifying  patterns of access related interventions I, the case of an EXIM}
Working with Bitguilder the following forms of capture performed by agent C in relation to address (public key) $k$ 
to which agent A had (or thought of having)  exclusive access 
(by way of exclusive access to the corresponding secret key $s$) can be distinguished:
\begin{description}
\item{\em Capture of access (CoA).} Capture of A's access to $k$ by C consists of a process at the end of which  the secret 
key $s$ for $k$ has become accessible to C possibly without the prior consent of A. After capture of access to $k$
 the corresponding secret key $s$ has become accessible to C.\footnote{%
In the terminology of \cite{BergstraL2013} after capture pseudo-monopresence of $s$ is lost indefinitely or
temporarily lost because of the capture by C.}

After secret key capture for $k$ the address $k$ is compromised. In addition the conjectural almost 
pseudomonopresence of
$s$ is broken (from the perspective of A) and a state of multipresence of $s$ is entered.
\item{\em CoA with subsequent coin capture (CoA+SCC).} 
CoA+SCC involves capture of access to $k$ followed by a transfer of an informational 
coin $q \leq q(k)$ from $k$ to an account,
say $r$ outside control of $A$.\footnote{For the definition of $q(k)$ we refer to the 
specification of the Nakamoto architecture in \cite{BergstraL2013}.}

After CoA to $k$ with SCC the address $k$ is 
compromised and an amount $q$ is lost (indefinitely or temporarily)  for A.

\item{\em Pseudo-theft.} Following \cite{BergstraL2013} Buitguilder ownership does not exist 
(because Buitguilder is supposed to qualify as an EXIM) and as a consequence theft
cannot be explained in terms of loss of control of an owner. Instead of theft we will define
pseudo-theft for an exclusively informational money. 

\item{\em Legality hypothesis for transfers.} A transfer may be illegal in the EXIM case if it constitutes part
of a larger scheme of illegal activity where the transaction plays the role of a payment. The legality hypothesis 
expresses the assumption that a distinction between legal transfers and illegal transfers can in 
principle be made.\footnote{%
If A makes an illegal transfer to B, B may be held accountable and blockchain inspection may provide the 
proof for law enforcing authorities that B's role was problematic. In the case of an EXIM, this state of affairs will not
render the amount received by B illegal (or black, or fraudulently acquired). The only role of the EXIM in this kind of case is keeping track of illegal behavior is that it may support linking B to an illegal transaction.}

\item{\em Legality hypothesis for CoA+SCC interventions.} We will assume that an activity of
CoA+SCC may be either classified as legal or as illegal. An illegal event of CoA+SCC at the expense of A  might involve extortion or otherwise unlawful ways of obtaining access to A's wallet password or access to A's 
physical storage of private keys for A's accounts.

Having available the legality distinction, illegal CoA+SCC may be considered a defining instance of pseudo-theft. 

\item{\em Incidentially legalized pseudo-theft.}
Legal cases of pseudo-theft are found for instance in the following case: 
(i) previously an amount is thought (by prosecution and/or police forces)  to have been
transferred to, or by, A illegally (this requires the legality hypothesis for transfers) and, (ii) CoA+SCC against A is effected by police forces under temporary legal cover of the prosecution, 
because (iii) after CoA+SCC the probability that
A can be linked to the mentioned illegal transfer is significantly higher, or (iv) in preparation of a penalty issued to
A which is effected by not returning part of the captured amount (in this case the penalty is measured against
the unlawfulness  of the transfer in qualitative terms, not in view of a flown analysis of the amount which is supposed
to reveal that ``it was stolen'', or illegally acquired).

\item{\em Denial of access (DoA).} Denial of access to $k$ takes place if 
A's access to $s$ is broken (indefinitely terminated, destroyed against the will of A). DoA takes place if all of A's
stored copies  of $s$ are deleted, or destroyed, or  irreversibly modified, or made inaccessible.

DoA splits in two cases:
\begin{description}
\item {\em Temporary DoA.} In this case obstacles of some form prevent A from having access to $k$ but a return
to the original state will (or at least can) take place at a future moment.

Temporary DoA is consistent with almost pseudomonopresence of $s$ and with multipresence of $s$.
\item {\em Indefinite DoA.} In this case A is not able to recover from its lack of access to A on its own.

In case of indefinite DoA absence (almost nonexistence) of $s$ may replace preceding 
almost pseudomonopresence.
\end{description}

After indefinite DoA the amount $q(k)$ may be lost altogether (even without SCC taking place). 
A can reduce there risk of DoA by having independent multiple storage of $s$.\footnote{%
The forms of storage may include memorization of a natural language encoding of $s$ 
according to some memorization scheme. In other words sometimes an amount can happen to
be stored in the brain of a single human participant only, with no other backup at hand.}
C cannot be sure that A has no backup copies of $s$.
Unfortunately, by A giving up pseudomonopresence of $s$ the risk of CoA (for $k$) increases.

\item{\em CoA and DoA (CoA+DoA) of  $k$.} CoA+DoA is 
performed by C if C captures access to $k$ and subsequently causes DoA for $k$ to A.

CoA+DoA leaves A in about the same position as CoA+SCC for an amount $q(k)$.

\item{\em Malicious linking (ML).} Malicious linking takes place if say C makes B 
believe that A is in control of $k$ while in fact C (or another agent friendly to C but unfriendly to A) is in control of $k$. 

Plausibly ML precedes an attempt by C to misguide B and have it transfer an amount to $k$ in 
order to perform a transaction with A. 
ML can be effective (for C) after CoA (by C) in circumstances where A fails to warn B that $k$ is 
not in A's exclusive control anymore. After CoA $s$ is not monopresent anymore, 
$k$ is compromised and A cannot be sure
to gain access to all amounts that are transferred by B to A.

\item{\em Malicious unlinking (MU).} Malicious unlinking (of A and $k$) takes place if say C makes B 
believe that A is not in control of $k$ while in fact A is in control of $k$ (and C is knowledgeable for that). 

MU may be attempted by C with the objective to move A into a state where it expects a transfer from B 
on an account, say $k^{\prime}$, which is less well-protected than $k$ against CoA or against DoA.

\item {\em Contaminating donation.} Suppose that B donates an amount $x$ to account $k$ in control of A,
and in such a way that A's control of $k$ has been acquired illegally. Now B's amount on $k$ becomes 
contaminated with amounts with an illegal background thus potentially creating legal problems for A. This may
be understood as an attack on A's integrity (as a user of account $k$). 

\item{\em Contaminating extortion.} Suppose that an anonymous peer B publishes the request that a
prospective victim of extortion, say V, must transfer an amount $x$ to an account $k$ in order to avoid the 
occurrence of
some action by B detrimental for V (the action lying probably outside the monetary system at hand). 
In this case (even if V refuses
to issue the required transfer to $k$) the account $k$ becomes contaminated with the public knowledge of
its use for this episode of (attempted) extortion. That contamination may create major difficulties for an agent A
who is in exclusive control of $k$ and who may just as well be a victim of B's actions as V is, in particular  if $B$'s
identity (as the agent being in control of $k$) has not yet been made public. 
This form of extortion almost forces B to reveal its identity in order not
to be linked in later stages with the episode of extortion, and as such it may be considered an attack against $k$
and (implicitly against A). We propose to call this a contaminating extortion attack. 
Contaminating extortion is a possible attack against
each account holding a non-negligible amount.

\end{description}
In principle A will try to maintain access to $k$ private from the moment that A has obtained 
control over $k$. A will operate in such a way that the following holds (or is done):
\begin{enumerate}
\item A has performed and continuously maintains some public announcement that it is in exclusive control of $k$. 
By doing so, A creates the belief in other agents
of A's adoption of conjectural almost pseudomonopresence of the secret key corresponding to $s$.
\item A systematically tries to prevent CoA of $k$. 
\item A makes, and safely stores, 
external copies of the secret key $s$ for $k$ in order to minimize the risk of DoA to $k$,
thus reducing (conjectural) pseudomonopresence to controlled multi-presence.
\item At no stage CoA of A's access to $k$ has occurred in the past. 
\item If, however, A becomes aware that CoA of $k$ has occurred, (and conjectural almost 
pseudomonopresence must be given up in favor of conjectural multi-presence),
A must (i) announce a denial of its link with $k$, 
and (ii) for an indefinite time watch out for ML attacks that use previous, but now outdated, 
announcements (by A)  of its control over $k$, and (iii) if still possible, secure its holdings in $k$ by transferral
to another account.
\item At no future stage DoA of A's access to $k$  will occur. 
\item Together the preceding items imply that A is justified in having belief in pseudomonopresence of $s$ (that is
in adopting conjectural pseudomonopresence for $k$).
\item ML of $k$ to an agent (say A$^{\prime}$) different from A is noticed and remedied.
\item ML of A to another public key (say $k^{\prime}$) different from $k$ and not in control of A  
(to the best of A's knowledge)  is noticed (by A) and remedied.
\item MU of $k$ from A is noticed and remedied.
\end{enumerate}

On the basis of this embryonic theory, terminology, and this survey of access related misuse 
or incidentally legalized use, some 
questions can be put forward. The questions that we suggest to pose about access related misuse are these.
\begin{enumerate}
\item  Is this classification of misuse  is sufficiently refined so that practical cases are covered? 
\item Are there important cases of misuse missing from this classification?
\item Taking the ``theory'' just presented as a point of departure: which conjectural abilities, in the sense 
of~\cite{BergstraDV2011}, can be put forward as a constituent of claims of relevance for this ``theory''.
\end{enumerate}
\subsubsection{Capture of access and denial of access for Bitpenny}
Assume a specification for Bitpenny (see Paragraph~\ref{bitpenny} above). 
Then the question is to develop a theory of use and misuse for Bitpenny.

\subsubsection{Classifying  patterns of access related interventions II, the case of Bitcoin}
In object oriented terms one may think of Bitcoin as a class extension of the EXIM Bitguilder. Bitguilder
is extended with the additional (non-EXIM) feature of ownership of amounts with the understanding that ownership 
may or may not coincide with haven access to an amount. In particular (in Bitcoin or ant non-EXIM) 
an agent may have access without ownership to a  stolen amount. Enriching an EXIM with non-EXIM features
a non-EXIM is obtained, for instance Bitcoin. Now additional aspects of legality enter the picture.
\begin{description}
\item {\em Patterns inherited from the EXIM case.} Patterns of use and capture for EXIMs inherit to the 
non-EXIM case of say Bitcoin. Moreover a concept of theft emerges, and additional cases of legal pseudo-theft 
arise.
\item {\em Theft (for a non-EXIM only).} In the case of Bitcoin or any other non-exclusively informational money, 
theft can be understood as illegal pseudo-theft. In an exclusively informational money (e.g. Bitguilder) theft 
does not exist. 
In Bitcoin for instance, theft (of $k$ from A) refers to
an illegal combination CoA+SCC, but not to a combination CoA+DoA.
\item {\em Legality hypothesis for amounts.} In a non-EXIM such as Bitcoin amounts may all or in part be
either legal or illegal depending on the legality of transactions (or events of capturing) by means of which 
an agent has acquired access to the amount.
\item {\em Legal pseudo-theft for illegally acquired amounts.} In there is a significant suspicion than control
over an amount has been illegally required by A, authorities may incidentally permit the police force to
perform an act of CoA+SCC against A thus capturing that very amount, or an equivalent amount.
\end{description}

\subsubsection{Legalized Bitcoin capture}
How to define legal forms of CoA+SCC in the case of Bitcoin? And how to define these legalized forms 
by viewing Bitcoin as ``an EXIM plus ownership".

\subsection{Transaction and mining related misuse}
It is not obvious that performing a double-spending attack comprises a misuse of Bitcoin. Neither is it obvious that a 
reverse
mining attack exclusively aimed at destroying a tail of the blockchain in preparation of transactions to the
advantage of the attacking miner constitutes a misuse. 

Intuitively that is the case, but from the perspective of exclusively
informational money (EXIM)  of \cite{BergstraL2013} such attacks are morally unproblematic. Thus for Bitguilder, i.e. 
Bitcoin cast as an EXIM, both attacks are unproblematic.

\subsubsection{Is performing a double-spending attack on Bitcoin morally problematic?}
Between Bitcoin and Bitguilder are many options for systems of rules of engagement, elements of what has been called the
AP sheaf (autonomy privacy sheaf) around Bitcoin, 
for which Bitguilder constitutes an extreme element. This leads to these questions:
\begin{enumerate}
\item Is performing a double-spending attack part of the rules of engagement for Bitcoin, or is it ethically wrong?
\item Assuming that performing double-spending attacks is considered a wrongdoing, on may consider a 
clone of Bitcoin, say BitcoinDS, less permissive than Bitguilder but permissive of double-spending attempt. Is BitcoinDS
a reasonable or even attractive option, or is its offering to its clients conceptually problematic?
\end{enumerate}
\subsubsection{Theory of forks}
Forks in Bitcoin result from double-spending attacks and from reverse mining attacks. Forks and fork recovery 
are at the heart of Bitcoin. It is unclear at this stage to what extent the very existence of 
forks constitutes an unavoidable feature of future informational monies. The question is to develop a theory of forks and to
assess to what extent forks are a fact of life for informational monies designed 
under the constraint of avoiding singe points of failure, or merely constitute a feature of Bitcoin's current implementation.

\section{General matters}
Under general matters we collect questions with a philosophical, legal, psychological, 
or ethical status, as well as questions posed from an evolutionary perspective.
\subsection{Bitcoin status}
\subsubsection{Informational money and philosophy of information}
Which aspects of the philosophy of information of \cite{Floridi2011} can be used to clarify 
or criticize the concept of informational money?
\subsubsection{Is Bitcoin a money.} Depending on one's views of monies and near-monies: 
is Bitcoin a money, or is it merely a near-money, or is it a commodity not qualifying as a near-money?
See also \cite{MaurerNS2013} for an account of the need to rethink true concept of 
money when contemplating systems like Bitcoin.
\subsubsection{Is Bitcoin an RPSF?}
In \cite{BM2011} it was indicated that disallowing interest payments can be viewed as 
removing a feature from a menu of features together constituting the mechanisms provided by a money. 
That leads to RPSF (reduced product set finance),
the connection with computing coming about when one notices that disallowing some conceivable 
features can make a system
or formalism more useful sometimes for unexpected and even deeply hidden reasons. 
As examples of profitable limitations of feature sets
taken from computing and logic one may consider:
\begin{itemize}
\item goto's in structured programming,
\item specialized instructions in a RISC architecture, 
\item global states in functional and logic programming, 
\item variables in propositional calculus, 
\item law of excluded middle in intuitionist reasoning, 
\item arbitrary comprehension in set theory.
\end{itemize}
Now as was suggested in \cite{BergstraL2013} Bitcoin seems not to provide the feature of a coin. Is that observation valid, 
and is it valid to the extent that a financial exchange system based on Bitcoin might be considered an RPSF for that reason?

\subsubsection{Other grounds for the classification of Bitcoin as an RPSF?}
The absence of ``true'' coins may not be considered grounds fro classifying Bitcoin as an RPSF, 
in which case the question can be posed if ether grounds for that classification exist.

\subsubsection{Which actions against Bitcoin participants are legal?}
The question about the legality of Bitcoin seems confounded with the presence of 
rather straightforward attempts to prevent
Bitcoin from gaining mileage by a coalition of established parties.\footnote{%
For an appraisal of regulation of Bitcoin see \cite{Kaplanov2012}.}
There is no doubt that survival of Bitcoin, if
that will occur at all, will be highlighted with a large number of minor and major legal issues that are raised with
the intent to discourage existing or prospective users.

The rights of established parties to dominate the use and the growth of Bitcoin and similar 
systems seems to be taken for granted by those who are in favor of established monies. 
We prefer to ask the question the following way: which actions against 
Bitcoin participants are legal? That depends on jurisdictions of course but general patterns can be investigated
independently from legal idiosyncrasies. 

\subsubsection{Are there signs of discrimination against Bitcoin?}
Bitcoin has reportedly been used for illegal purposes, but so have many other financial systems in existence today. 
It should not be the case that this unfortunate fact legalizes a general pattern of 
discrimination against those who consider Bitcoin 
an important development worth of real life experimentation. Are there signs of such discrimination?
\subsubsection{Demarcating ownership and access in an exclusively informational money}
There is a vast literature about ownership of information, see e.g.~\cite{Liu2001}. 
In an exclusively informational money 
(EXIM see \cite{BergstraL2013}) access replaces ownership as a central concept. 
But in a system built around an EXIM
ownership will be more important than access when it comes to more conventional, 
and for that reason more peripheral, parts of the system. For instance
someone may claim to own a self-generated public key (in excess of having access to it). 
The question is to determine where
in an EXIM the concept of ownership persists, and where the demarcation line between access and ownership 
is located?
\subsection{Psychology of money}
\subsubsection{Ultimatum games in Bitcoin: just like other money or not?}
Many psychological experiments have been conducted with simple games such as the 
ultimatum game and the dictator game, 
see for instance~\cite{BoltonZ1995}. A remarkable outcome of that work has been 
that human participants deviate from purely 
game theoretic predictions made on the basis of seemingly evident assumptions about participant preferences.

One can replay many psychological games in a setting of Bitcoin and try to see if replacing conventional money by Bitcoin
changes the situation. So in spite of having many features of a money Bitcoin might show measurable 
psychological differences from conventional monies that cannot be explained from a lack of familiarity. 
It might be the case that norms and expectations shift once the transition to a 
Bitcoin setting is made by subjects participating in an experiment.

\subsection{Rationale of Bitguilder}
\subsubsection{Which thought experiments concerning Bitcoin are permissible?}
As was mentioned above, in \cite{BergstraL2013} we have introduced 
Bitguilder as a (hypothetical) informational money that technically 
works like Bitcoin but which offers is participants different rules of engagement. In particular in Bitguilder 
access to an amount
takes priority (in Bitguilder) over ownership. As a consequence amounts of BGU (Bitguilder unit) cannot 
be stolen, but can only be captured. Such amounts cannot be under control of an agent in such a way that the
acquisition thereof constitutes a misuse of the Bitguilder system. In other words 
Bitguilder amounts can be acquired in return
for illegal actions services or goods, but not technically in illegal or incorrect ways.\footnote{%
The whole story of Bitguilder is more involved and has been captured in the concept 
of an exclusively informational money (EXIM) in \cite{BergstraL2013}.}

Our reason for renaming Bitcoin into Bitguilder in preparation of this thought experiment has been that we do not want
to state in any form that in the case of Bitcoin theft is theoretically impossible 
(and for that reason admissible because access is all that matters). We see
no way to carry out that thought experiment about Bitcoin without making suggestions 
that can be misunderstood (and that for that very reason should not be made). Have we been overly cautious 
and could the thought experiment concerning an EXIM that makes use of
Bitguilder have been properly performed without the introduction of a new term, that is in term of Bitcoin?

\subsubsection{Is there a special field of Bitcoin ethics?}
Bitcoin usage induces several ethical questions, both at design level and at the level of users. 
 The question is to what extent novel ethical principles in connection with Bitcoin will emerge.
 
\subsection{Bitcoin and the evolution of informational monies}
In \cite{BergstraL2013} a method has been outlined to use the concept of a portfolio of natural kinds for establishing the 
role that Bitcoin as a system may play in the evolution of informational money. That role is qualified  in terms of the update that 
the appearance of Bitcoin (comparable to a genetic mutation of the stock of available programs)
may imply for the
portfolio of natural kinds that makes up for the ``genetic code'' of informational monies. 

\subsubsection{Relevance for Bitcoin of a natural kind portfolio based approach}

The questions that can be posed about the natural kinds portfolio based approach are many, for instance:
\begin{enumerate}
\item Is the application of natural kinds as a method for the quantification and qualification of artifact evolution justified?
\item Is the approach through natural kind portfolio updates applicable to software evolution? An more specifically to Bitcoin?
\item Is an assessment of the role of Bitcoin in the evolution of informational monies from the perspective of 
software evolution an adequate approach or must a significantly more cognitive and less mechanical approach to this
question be preferred?
\end{enumerate}

\subsection{Interest prohibition and gambling prohibition}
Some financial systems maintain interest prohibition and gambling prohibition as ethical norms.\footnote{%
For instance Islamic finance (see e.g. \cite{BM2012}) may be seen as a family of financial systems each of 
which must comply with the following five 
normative constraints:  IP: interest prohibition, GP: gambling prohibition,  PoMT: prohibition of misleading transactions,
PSNE: prohibition of selling  non-existing (including unfinished)  entities, and ORD($p,e$): 
obliged regular donations (a fraction $p$ of one's possessions) to  one or more ethically approved destinations
(with ethical approval granted in accordance with some criterion $e$ that serves as a parameter to this obligation).}
That motivates the following questions.
\subsubsection{Can Bitguilder support the implementation of interest free finance?}
In \cite{BergstraL2013} we have put forward the suggestion that for Bitguilder debt does not exist and
for that reason interest payment is out of the question. That leads to contemplating a dual system where a copy of 
a conventional money (say EUR), now understood as a near-money and explicitly not as a money, 
is combined with a Bitguilder system. 
In this combined system interest payments in EUR are considered 
unproblematic because of its non-money status. The justification for assigning to EUR a non-money 
status are found in its low and politically always vulnerable quality as a means of storage of value. Given a certain
level of inflation, interest payments cannot be understood as an increase of a real amount 
but merely as a moderation of a steady decrease of accumulated value. 

This system might be complemented with a third layer ``below'' EUR that implements demurrage as 
proposed by Gesell.\footnote{Freicoin is a Bitcoin-like system that implements demurrage, see \cite{Steadman2013}.}
Demurrage is a forceful devaluation of money-items in the hands of their owner.
Demurrage is meant to function as an incentive against hoarding and its effect can be enhanced by restricting the
money-items to local use thus acting against globalization as well. Money that deprecates predictably 
by way of demurrage has become known in the USA under the term stamped money.

In a three-level system EUR is sandwiched between two innovations: 
Bitguilder (as a hypothetical legal casting of Bitcoin) and stamped money.

The question that we suggest to be posed is to what extent these considerations make sense. 
For instance one might object that 
interest prohibition is equally negative about non-monies and is insensitive for a decrease in ``real value'', 
being somehow intrinsically linked with nominal measurements.

\subsubsection{Is Bitcoin mining based on gambling?}
Bitcoin mining seems to make use of a mix of gambling and probabilistic programming. 

A user, say $u$, of a dedicated 
Bitcoin mining machine\footnote{%
Bitcoin mining machines are in fact constrained (not all inputs are allowed) first set-preimage 
(element of the preimage of a set) attack generators for SHA-256 with 
more or less programmable strategies.}
(other prospective miners outside a mining pool 
have little chance of mining success anyhow) must produce (or at least define)  
a sequence $w^u_n, n \in \Nat$ of words (say bit sequences of length 512 extending a given 256 bit sequence $w_i$)
and search for the first element in the sequence on which the secure hash function SFA-256 
(see \cite{FIPS2008}) produces a result (a bit string
of length 256) starting with some given number (not exceeding 256) of initial 0's. 
We recall the details of the simplified abstraction of Bitcoin mining put forward in
Question \ref{complexitySHA} and write SHA-256-512 for the restriction
of SHA-256 to inputs of length 512.

Although SHA-256 has a clear and algorithmic definition, producing its output is like throwing 
a dice in that you cannot predict the output with any reasonable probability above $256^{-1}$ without 
actually computing it. 
That there are $k$ initial zeroes has a probability of about $2^{-k}$.

The owner of the fastest Bitcoin miner may use a straightforward enumeration in increasing 
order  to define an enumeration $w^u_n$ implicitly defined by  
$\texttt{bs2n}(w^u_n) =\texttt{bs2n}(w_i)  \times 2^{256}  + n $  with $w_i$ a 256 bit input string. 
For an owner $u^{\prime}$ of a slower 
Bitcoin mining machine employing the same 
strategy (word generator) is useless and by way of gamble another strategy must be chosen. 
Another strategy may work as follows for instance: starting with $\texttt{bs2n}(w_i)$ and then 
repeatedly adding some prime $p$ (modulo $2^{256}$)
with  prime $p < 2^{256}$ privately generated  and kept secret w.r.t. other members of the mining community.

It seems that most miners need to make a gamble concerning their enumeration strategy
of bit sequences on which SHA-256-512 is to be performed and to keep the result of that gamble secret. 

It is an algorithmic question to find out to what extent this
analysis of Bitcoin mining is valid. It is a non-trivial issue concerning the concept of gambling 
prohibition to find out to what extent this story
might have an impact on the ethical status of Bitcoin mining (and Bitcoin use dependent on mining for validation) 
from the perspective of gambling prohibition.

Assuming that Bitcoin mining would be found defective from the perspective of gambling prohibition, 
then it is an algorithmic problem to find a different
system for rewarding transaction validation in a P2P network. 
That mechanism is probably the main finding of Nakamoto, conceivably with the status of a 
conceptual breakthrough in the area of informational money.

\section{History of informational money}
Many discussion of money use the history of money, or conjectures about that history, as an essential explanatory tool. 
Historic explanations abound in the context of conventional monies. We expect that it will be the same with 
informational monies.

Assuming that informational money is the future of money in some form or another, it is efficient that its history
be written in real time so that in contrast with conventional monies conjectural histories of informational 
money won't be needed in the future. For that reason we think that ``historic'' questions about informational monies are 
far more important than for average distributed software functionalities.
\subsection{Technical history of informational money}
\subsubsection{What has been the role of digital cash in the development of Bitcoin?}
Bitcoin has now appeared with such force in the media that one might think that informational monies
are novel. But that is not true. The primary pioneer of informational money is David Chaum.
 David Chaum's digital cash has paved the way for Bitcoin and it might still
turn out to be of more lasting importance. A historic question that might be posed however is this: what has the impact
been of Chaum's work on the design of Bitcoin. 

Clearly this question has become rather hairy because the designer(s)
of Bitcoin prefer an unusual form of anonymity inspire of the success that has already been achieved, but
we feel that the question is not ill-posed for that reason.

\subsubsection{Chaum's role in connection with the evolution of fiat monies}
It would be a misrepresentation of Chaum's work to portray it as being of relevance to the development
of informational monies only. A question that we consider to be of independent historic interest is 
which impact Chaum's work has had (and is still  likely to have) on the 
(future) evolution of conventional fiat monies.

\subsubsection{Separating the  history of informational money from that of information security}
Both Chaum's developments leading to digital cash and Nakamoto's plan for Bitcoin put cryptography 
and information security central stage to such an extent that informational money seems to be merely 
a topic in information security. Can the histories of these subjects be separated: more specifically, can a history
of informational money be given that is not dominated by aspects of information security?\footnote{%
For the history of information security we refer to \cite{Leeuw2007,LeeuwB2007} and references mentioned
 there.}

\subsection{Bitcoin and the history of finance}
\subsubsection{It is reasonable to consider Gesell, Maududi, and Nakamoto as somehow related?}
In \cite{BergstraL2013} we have put forward that Gesell, Maududi, and Nakamoto may be 
viewed as financial revolutionaries with opposing but still related views, each having  their own
vision of the impact of the (development of the) quantity of money. These three names share a noticeable  and
persisting impact on non-mainstream monies and systems: respectively 
LETS (local exchange and trading systems), IF (Islamic finance), and Bitcoin. \footnote{%
We have not included Chaum in this list because, in spite of his influence on the development of 
informational money and on the interactions
between informational money and information security,
his work seems not to aim at changing the ideology behind money itself.}

Is this combination of non-mainstream financial and monetary thinkers meaningful, or should other combinations of 
names be contemplated. In particular it may be considered misguided to connect Nakamoto to Gesell and Maududi, 
while connections with other innovative authors on monies would have been more illuminating?

\section{Economic questions}
Economic questions about Bitcoin in particular and about informational monies in general concern functionality or potential functionality
in the world at large, that is, how a system serves the needs for which it has been designed or for which it has become a principal tool.
Economic questions on informational monies are considered under the assumption that as as pieces of information technology
the systems are adequate.
\subsection{Sustainability of Bitcoin}
\subsubsection{Are small changes in the Bitcoin system relevant for its survival?}
If Bitcoin is viewed as merely a computer game striving for a dominant market position, then minor modifications in
the software may decide upon its fate, because survival is a matter of winning some beauty contest. Is it necessary for the
persistence of Bitcoin to apply small changes to its protocol?

\subsubsection{Is survival of Bitcoin dependent on coexistence with other informational monies?}
Can Bitcoin can  persist otherwise than in in coexistence (symbiosis) with other and different informational monies.

\subsubsection{Vending machine usage}
Is the possible use of Bitcoin for vending machines (see~\cite{BamertDEWW2013}) of any relevance for its survival?
\subsection{Eventual demise of Bitcoin}
It seems implausible that Bitcoin will exist forever, and for that reason it must be expected that sooner or later 
Bitcoin will come to an end. This assumption by itself leads to questions. In~\cite{Hanley2013} a wide spectrum of arguments is listed all supposedly pointing towards the expected demise of Bitcoin.
\subsubsection{How will Bitcoin disappear (if at all)?}
When Bitcoin disappears that may work in different ways. First of all it is not clear under which conditions Bitcoin must or can be
said to be ``out of use''. Perhaps a rigorous criterion for that state of affairs must be developed? 
Such criteria are required if simple questions like ``how many informational monies 
are in existence'' are to have well-defined answers.
\subsubsection{Can an informational money have a predetermined life-cycle?}
Uncertainties about the demise of an informational money can be removed by making its termination part of the design
from the start. That
is by taking the final stages of its life-cycle explicitly on board as being in need of requirements engineering and designing
protocols and software in such a way that the final stages of its existence comply with the resulting requirements. The question 
becomes: how can a modified Bitcoin including a termination phase look like.

\subsubsection{Scenarios for Bitcoin's demise}
Is it conceivable that Bitcoin comes to an end while the external value of a BTC is still significant? 

If not, is the
moment that the external value of a BTC is zero EUR exactly the moment that Bitcoin has stopped being an existing
informational money (or near-money)?

Is it conceivable that Bitcoin remains in existence for a very long time
 as a well-preserved
legacy with BTCs having positive value, but not in use for any practical purpose, comparable to ancient gold coins,
which have value without having the status of money? Is this scenario more likely if Bitcoin loses out against one
of its many derived but competing developments?

In other words: can the demise of Bitcoin take the form that
a quantity of BTC becomes an informational commodity with historic value, the value connected to its historic but from
some moment onwards deprecated role as an informational coin?

\subsection{Ultimate perspective of Bitcoin}
Ignoring Bitcoin's demise one faces question concerning the importance that it might eventually acquire.
\subsubsection{External Bitcoin valuation: what can be done?}
In \cite{BergstraL2013} we have estimated the value of a BTC at 50 EUR. 
We believe that value to be safe on the low side, that is 1 BTC is defensibly valued more than 50 EUR.

Since the writing of \cite{BergstraL2013} the value in USD if Bitcoin has gone up to 1200 and down again to 600.
Are such fluctuations confirmations of a claim that a BTC is worth at least 50 EUR or are such fluctuations immaterial for the assessment of such claims?

The more general question arises whether this kind of estimate makes any sense, and if so how it can be 
improved and how it must be maintained. If not, are other means to estimate a lower bound for the 
BTC to EUR ratio available?

\subsubsection{EUR volatility: how to explain?}
The BTC to EUR ratio turns out to be fairly fluctuating. From the perspective of a Bitcoin participant that 
signifies high volatility of the Euro. What is it in the management of the Euro, or in the Eurozone economy, or in Eurozone 
policies that may explain this volatility? 

\subsubsection{Bitcoin deflation and stabilization: can intermediate parties be helpful?}
Can BTC steadily (though with ups and downs) increase in value against EUR and at the same time become
an important means of exchange? A continuous change in value (not only against EUR but also and more importantly against 
a bundle of goods and services) induces high transaction costs on parties that want to use BTCs 
as a means of exchange rather than as a store of value.

Can Bitcoin based services, perhaps to be provided by new intermediate parties, 
be developed that cushion its users against BTC purchasing power
fluctuations in the light of a steady but still unpredictable increase in value of BTC versus EUR.

\subsubsection{Can the Euro become Bitcoin backed?}
Rather than asking to what extent Bitcoin can safeguard its value by becoming EUR backed, 
we prefer to ask if the opposite connection can come about. Is it conceivable that the monetary system evolves in 
such a way that BTC backs the EUR, like gold might have done in another world?

\subsubsection{Will ordinary coins become outdated?}
In~\cite{Zaag2012} the question is raised if ordinary coins and banknotes will have a long future simply because
of their independence from network technology, power supply and so on. If this question has a positive answer, that implies that
a Bitcoin monoculture is unlikely to be in place soon, and not even desirable. The question is to what extent IT independent means
of exchange, which must unavoidably be limited to non-informational money-items, will be needed in the future?
\subsubsection{Is Bitcoin a risk for the conventional financial system?}
Some consider Bitcoin to constitute a potentially disruptive technology; indeed
In many ways Bitcoin might be a challenge of conventional finance. 
The question is to provide a convincing risk analysis for this matter from the perspective of the maintenance of conventional monies.

 {\bf Acknowledgement.} Discussions with Andrea Haker and Sanne Nolst Trenit\'{e} (both University of Amsterdam) 
 have contributed to this paper.
 
\bibliographystyle{plain}

\addcontentsline{toc}{section}{References}
\begin{thebibliography}{58}

\bibitem{AndroulakiKRSC2012}
Elli Androulaki, Ghassan O. Karame, Marc Roeschlin, Tobias Scherer, and Srdjan Capkun. 
\newblock Evaluating user privacy in Bitcoin.
\newblock \texttt{in: Proc. 2012 ACM Conf. on Computer and Communications Security} (2012).

\bibitem{BabaioffDOZ2012}
Moshe Babaioff, Shahar Dobzinski, Sigal Oren and Aviv Zohar.
\newblock On Bitcoin and red balloons.
\newblock {\em Proc EC'12 Valencia, Spain,} (2012).

\bibitem{BamertDEWW2013}
T.\ Bamert, C.\ Decker, L.\ Elsen, R.\ Wattenhofer, and S.\ Welten
\newblock Have snack, pay with Bitcoins.
\newblock {\em 13th International Conference of Peer-to-Peer Networks,} (2013).

\bibitem{Barber2012}
Simon Barber, Xavier Boyen, Elain Shi, and Ersin Uzun. 
\newblock Bitter to better--how to make Bitcoin a better currency.
\newblock {\em In: A.D. Keromytis (ed.): FC 2012}, LNCS 7397, 399--414 (2012).

\bibitem{Berciu2013}
Radu Mihai Berciu.
\newblock Designing Incentives in P2P Systems.
\newblock {\em MSc Thesis, Baylor University,} 
\texttt{https://beardocs.baylor.edu:8443/xmlui/handle/2104/8811} (2013).

\bibitem{Bergstra2012}
Jan A. Bergstra. 
\newblock Decision taking for selling thread startup.
\newblock \texttt{arXiv:1208.2460 [cs.SE]} (2012).

\bibitem{Bergstra2012d}
Jan A. Bergstra. 
\newblock Informaticology: combining computer science, data science, and fiction science.
\newblock \texttt{arXiv:1210.6636 [cs.SE]} (2012).

\bibitem{Bergstra2013a}
Jan A. Bergstra. 
\newblock Formaleuros, formalbitcoins, and virtual monies.
\newblock \texttt{arxiv.org/abs/ 1008.0616v2 [cs.CY]} (2013).

\bibitem{BergstraB2008}
Jan A. Bergstra and Mark Burgess.
\newblock A static theory of promises.
\newblock \texttt{arXiv:0810.3294 [cs.MA]} (2008).

\bibitem{BBP2013}
Jan A. Bergstra, Inge Bethke, and Alban Ponse.
\newblock Cancellation meadows: a generic basis theorem and some applications.
\newblock {\em The Computer Journal}, 56(1):3--14, \texttt{doi:10.1093/comjnl/bsx147} (2013).

\bibitem{BergstraDV2011}
J.A. Bergstra, G.P.A.J. Delen and S.F.M. van Vlijmen. 
\newblock Outsourcing Competence.
\newblock \texttt{arXiv:1109.6536 [cs.OH]} (2011).

\bibitem{BergstraL2013}
Jan A. Bergstra and Karl de Leeuw. 
\newblock Bitcoin and Beyond: Exclusively Informational Money.
\newblock \texttt{arXiv:1304.4758v2 [cs.CY]} (2013).

\bibitem{BL02}
J.A. Bergstra and M.E. Loots.
\newblock Program algebra for sequential code.
\newblock {\em Journal of Logic and Algebraic Programming},
51(2):125--156 (2002).

\bibitem{BM07}
J.A. Bergstra and C.A. Middelburg. 
Thread algebra for strategic interleaving. 
\newblock \emph{Formal Aspects of Computing}, 19(4):445--474 (2007).

\bibitem{BM2009}
J.A. Bergstra and C.A. Middelburg. 
\newblock Machine structure oriented control code logic.
\newblock \emph{Acta Informatica}, 46(5): 375--401 (2009).



\bibitem{BM2011}
J.A. Bergstra and C.A. Middelburg. 
\newblock Preliminaries to an investigation of reduced product set finance. 
\newblock \emph{JKAU: Islamic Economics}, 24(1):175--210 (2011).

\bibitem{BM2012}
J.A. Bergstra and C.A. Middelburg. 
\newblock Interest prohibition and financial product innovation. 
\newblock \emph{In: Finance Islamique: Regard(s) sur une Finance Alternative, 
Mazars Hadj Ali}, 274--284 (2012).

\bibitem{BergstraM2013}
J.\ A.\ Bergstra and C.\ A.\ Middelburg.
\newblock Instruction sequence expressions for the secure hash algorithm SHA-256.
\newblock \texttt{http://arxiv.org/abs/1301.3297 [cs.PL]}, (2013).

\bibitem{BergstraM2013b}
J.\ A.\ Bergstra and C.\ A.\ Middelburg.
\newblock Long multiplication by instruction sequences with backward jump instructions.
\newblock \texttt{http://arxiv.org/abs/1312.1812v2 [cs.PL]}, (2013).

\bibitem{Boehme2013}
Rainer B\"{o}hme.
\newblock Internet Protocol Adoption: Leaning from Bitcoin.\\
\newblock \texttt{http://www.iab.org.p.loopop.net/wp-content/IAB-uploads/2013/06/itat-\\2013\_submission\_17.pdf},
Universit\"{a}t M\"{u}nster, (2013).


\bibitem{BoltonZ1995}
Gary E. Bolton and Rami Zwick.
\newblock Anonymity versus punishment in ultimatum game bargaining.
\newblock {\em Games and Economic Behavior,} 10: 95--121 (1995).

\bibitem{BosEA2013}
J.\ W.\ Bos, J.\ A.\ Halderman, N.\  Heninger, J.\ Moore, M.\ Naehrig, and E.\ Wustrow.
\newblock  Elliptic Curve Cryptography in Practice. 
\newblock Microsoft Research. November (2013).

\bibitem{BritoC2013}
Jerry Brito and Andrea Castillo.
\newblock Bitcoin, a Primer for Policymakers.
\newblock {\em Mercatus Center, George Mason University,} (2013).

\bibitem{Bruce2013}
J.\ D.\ Bruce.
\newblock Purely P2P Crypto-Currency With Finite Mini-Blockchain.
\newblock \texttt{http://www.bitfreak.info/files/pp2p-ccmbc-rev1.pdf}, (May 2013).

\bibitem{CourtoisGN2013}
Nicolas T.\ Courtois, Marek Grajek, and Rahul Naik. 
\newblock The Unreasonable Fundamental Incertitudes Behind Bitcoin Mining.
\newblock arXiv preprint \texttt{arXiv:1310.7935}, (2013).
 
\bibitem{Entrup2013}
Gerion Entrup.
\newblock Bitcoin, Der St\"{a}rkere gewinnt.
\newblock {\em Thesis Leibniz Universit\"{a}t Hannover, Institut f\"{u}r Theoretische Informatik,}\\
\texttt{http://www.thi.uni-hannover.de/fileadmin/forschung/arbeiten/entrup-ba.pdf}
(September 2013).

\bibitem{FIPS2008}
FIPS 180-3.
\newblock Secure Hash Standard. 
\newblock \emph{NIST} (2008).

\bibitem{Floridi2011}
Luciano Floridi.
\newblock Philosophy of Information. 
\newblock {\em Oxford University Press}, ISBN 978-0-19-923239-0 (2011).


\bibitem{HalpernO2005}
Joseph Y. Halpern and Kevin R. O'Neill.
\newblock Anonymity and information hiding in multiagent systems.
\newblock {\em Journal of Computer Security,} 13: 483--514 (2005). 

\bibitem{Hanley2013}
Brian P.\ Hanley.  
\newblock The False Premises and Promises of Bitcoin.
\newblock arXiv preprint \texttt{arXiv:1312.2048} (2013).

\bibitem{Herrmann2012}
Matthias Herrmann. 
\newblock Implementation, evaluation, and detection of a double-spend attack on Bitcoin. 
\newblock \emph{MSc Thesis, ETH Z\"{u}rich} (2012).

\bibitem{Huang2013}
Danny Yuxing Huang.
\newblock Profit-driven abuses of virtual currencies.
\newblock \texttt{http://sysnet.ucsd.edu/~dhuang/pmwiki/uploads/Main/huang-research-exam.pdf}
UCSD, (2013).

\bibitem{HughesS2004} 
Dominic Hughes and Vitaly Shmatikov.
\newblock Information hiding, anonymity and privacy: a modular approach.
\newblock {\em Journal of Computer Security,} 12: 3--36 (2004).

\bibitem{JohnsonMV2001}
Don Johnson, Alfred Menezes, and Scott Vanstone. 
\newblock The elliptic curve digital signature algorithm (ECDSA). 
\newblock {\em IJCS}, 1,36--63 (2001).

\bibitem{Kaplanov2012}
Nikolei M. Kaplanov.
\newblock Nerdy money: Bitcoin, the private digital currency, and the case against its regulation.
\newblock {\em Temple University Legal Studies Research Paper} \url{http://ssrn.com/abstract=2115203} (2012)

\bibitem{Leeuw2007}
Karl de Leeuw. 
\newblock Introduction. 
\newblock \emph{In:~\cite{LeeuwB2007}}, 1--25 (2007).

\bibitem{LeeuwB2007}
Karl de Leeuw and Jan Bergstra (eds.). 
\newblock The history of information security--A comprehensive handbook. 
\newblock \emph{Elsevier} (2007).

\bibitem{Liu2001}
Joseph P. Liu.
\newblock Owning digital copies: copyright law and the incidents of ownership.
\newblock {\em William and Mary Law Review,} (42) 1245--1317 (2001).

\bibitem{MaurerNS2013}
Bill Maurer, Taylor C. Nelms, and Lana Swartz.
\newblock ``When perhaps the real problem is itself!'': the practical materiality of Bitcoin.
\newblock {\em Social Semiotics}, DOI:10.1080/10350330.2013.777594 (2013).

\bibitem{Nakamoto2008}
Satoshi Nakamoto.
\newblock Bitcoin: a peer-to-peer electronic cash system.
\newblock \url{http://Bitcoin.org/Bitcoin.pdf} (2008).

\bibitem{PfitzmannK2001}
Andreas Pfitzmann and Marit K\"{o}hntopp.
\newblock Anonymity, unobservability, and pseudonymity--a proposal for terminology.
\newblock{\em in H. Federath (Ed.): Anonymity 2000} Springer LNCS 2000, 1--9 (2001).

\bibitem{Steadman2013}
Ian Steadman. 
\newblock Wary of Bitcoin? A guide to some other cryptocurrencies. 
\newblock \url{http://www.wired.co.uk/news/archive/2013-05/7/alternative-cryptocurrencies}\\
\texttt{-guide/page/4} (2013).

\bibitem{ThalerA2008}
D.\ Thaler and B.\ Adoba.
\newblock What makes for a successful protocol?
\newblock RFC 5218, (2103).

\bibitem{Zaag2012}
Marijn van der Zaag.
\newblock Digitalisering van contant geld (in Dutch).
\newblock {\em BSc Thesis, Radboud University Nijmegen,} (2012).

\end{thebibliography}

\end{document}